\documentclass[Letter]{aa}
\usepackage{newtxtext,newtxmath}
\usepackage[colorlinks,citecolor=blue]{hyperref}

\usepackage[T1]{fontenc}
\usepackage{ae,aecompl}
\usepackage{float}
\usepackage{lineno}

\usepackage{graphicx}	
\usepackage{amsmath}	

\usepackage{bm}	        

\begin{document}

\title{Peanut/X-shaped bulges: Returning to the vertical resonance and separatrix}

\author{
Sergei A. Rusakov\inst{1}
\and
Viktor D. Zozulia\inst{1}
\and
Natalia Ya. Sotnikova\inst{1}
\and
Anton A. Smirnov\inst{1, 2}
}
\institute{
  St. Petersburg State University,
  Universitetskij pr.~28, 198504 St. Petersburg, Stary Peterhof, Russia \\
  \email{n.sotnikova@spbu.ru}
  \and
  Central (Pulkovo) Astronomical Observatory of RAS, Pulkovskoye Chaussee 65/1, 196140 St. Petersburg, Russia
}


\date{Received XXX; accepted YYY}





\abstract
{
We investigate the vertical structure and evolution of an $N$-body galactic bar by analyzing pure resonant orbits that form the skeleton of boxy/peanut/X-shaped (BP/X) bulges in Cartesian and action-angle spaces ($\theta_\mathrm{res}=\theta_z - \theta_R, J_z$). We show that the main backbone of this skeleton is made up of separatrix orbits associated with unstable figure-of-eight resonant orbits. 
For the first time, we examine the long-term behavior of the separatrix, showing how it sinks to the bottom of phase space and shrinks. As a result, ``fresh'' orbits involved in a flat bar are captured into vertical resonance, passing through it near the separatrix and creating orbital crowding. This crowding traces the thickest part of the BP/X bulge and outlines the ends of its X-structure rays. The rise of orbits leads to changes in the potential and geometry of phase space in these regions. Over time, when the rise ceases, evolution slows and stops.
}

\maketitle
\nolinenumbers



\section{Introduction}
\label{sec:intro}

Disk galaxies, when observed edge-on, often exhibit boxy/peanut/X-shaped (BP/X) bulges \citep{Lutticke_etal2000,Yoshino_Yamauchi2015,Marchuk2022} that protrude from the galactic mid-plane. It is generally accepted that such structures are the result of the secular evolution of bars in a vertical direction \citep{Combes_Sanders1981,Combes_etal1990,Pfenniger_Friedli1991,Raha_etal1991}. As for the flat parts of a bar, there is a main orbital family (x1) that forms its backbone \citep{Contopoulos_Papayannopoulos1980}. Thick parts are populated by orbits from the tree of its 3D bifurcating families \citep{Skokos_etal2002a}, as well as with periodic orbits of multiplicity higher than one \citep{Patsis_Athanassoula2019,Wang_etal2022}. 
The characteristic BP/X shape of the bar's vertical structure is created by various types of periodic orbits and there is an established understanding of the leading mechanism that causes orbits to elevate from the mid-plane associated with the vertical internal Lindblad resonance (vILR). This can lead to either resonant orbital capture or resonant heating (\citealp{Combes_etal1990,Quillen2002,Quillen_etal2014,Zozulia_etal2024b,Zozulia_etal2025}; see, however, \citealp{Sellwood_Gerhard2020}).
\par
The structure of the BP/X bulges can be analyzed not only using families of periodic orbits, but also based on the orbital
frequency ratio, $\omega_z/\omega_x$, where $\omega_z$ is the frequency of vertical oscillations and $\omega_x$ is the frequency along the $x$-axis in the frame rotating with the pattern speed, $\Omega_\mathrm{p}$ (e.g., \citealp{Portail_etal2015b,Abbott_etal2017,Parul_etal2020}). Frequency analysis shows that the BP/X bulge structure is layered, as demonstrated in previous works. The orbits in the central part of the BP/X bulge have smaller Jacobi integral values, $H_j$, and $\omega_z/\omega_x$ ratios than the orbits in its peripheral regions. In the outermost region, near the end of the BP/X bulge, $\omega_z/\omega_x$ is close to 2:1, which means that the orbits are predominantly banana-shaped (x1v1 orbits). BP/X bulges possess one prominent and bright feature, the so-called X-structure. \citet{Portail_etal2015b} proposed that the resonant boxlet orbits associated with the 
$\omega_z/\omega_x=5/3$ resonant orbits (brezel orbits) are responsible for the X-shape in the inner parts of a bar, although such orbits are very rare \citep{Abbott_etal2017}. 
\citet{Parul_etal2020} found that this structure in the unsharp masked images is outlined by the $z$-maxima of all orbits from which the BP/X bulge is assembled. This conclusion is consistent with an earlier result by \citet{Patsis_Katsanikas2014a}.
\par
In \citet{Portail_etal2015b,Abbott_etal2017} and \citet{Parul_etal2020}, an orbital analysis was carried out using single snapshots in systems in which orbital reconfiguration had already occurred. 
Orbits that join the bar during its evolution undergo a gradual transformation in the vertical direction \citep{Lokas2019,Sellwood_Gerhard2020,Li_etal2023,Zozulia_etal2024b,Zozulia_etal2025,Lokas2025}, and analysis of the orbital composition of snapshots reveals little about how these changes occurred.
\par
In this Letter we return to the question of BP/X bulge shaping by analyzing how the phase space of the $N$-body galaxy model is transformed over the long-term evolution of the bar. We describe the phase space using ``action-angle'' variables $(J_\mathrm{R},\,L_z ,\,J_z)-(\theta_\mathrm{R},\,\theta_\phi,\,\theta_z)$ calculated directly in the evolving $N$-body potential of the model \citep{Zozulia_etal2024a,Zozulia_etal2024b,Zozulia_etal2025}. The action-angle formalism allows one to investigate and visualize the evolution of the orbital skeleton, complementing the common approach based on four-dimensional Poincar\'{e} surfaces of section, which are difficult to visualize. The application of this approach to numerical models is thoroughly discussed in~\cite{Zozulia_2026}. Our analysis combines the  evolutionary approach by \citet{Quillen_etal2014} with an approach based on periodic orbits, primarily the x1v1 and x1v2 orbits. We classify all the bar orbits on the basis of the behavior of the resonant angle, $\theta_\mathrm{res}=\theta_z - \theta_\mathrm{R}$, for the entire possible range of $H_j$ at different time moments.
Such a classification allows us to divide phase space $J_z$--$\theta_\mathrm{res}$ into clearly distinguishable regions with different types of orbits. We distinguish between orbits located in the region bounded by the separatrix associated with the x1v1 and x1v2 orbits, and orbits lying outside the separatrix. The former orbits librate around fixed points (x1v1 orbits). The latter are circulating orbits. Circulating orbits with a decreasing resonant angle and large values of $J_z$ include some strictly periodic orbits (brezels and fish). We revisit the role of x1v1 orbits in the orbital transformation and formation of the peanut-shaped contour of the BP/X bulge and show for the first time directly for an evolving $N$-body model that the thickest part of the BP/X bulge and the outermost part of the X-structure are associated with a resonant separatrix for these orbits as predicted by \citet{Quillen_etal2014}.
\par
In Section~\ref{sec:model}, we describe our numerical model. 
In Section~\ref{sec:orbits},
we determine the positions (vertical action $J_z$ or $J_z^\mathrm{max}$) for x1v1 orbits (fixed points), separatrix orbits, and some strictly periodic orbits, brezels and fish. 
Then we construct from these orbital bones the phase skeleton of our model and track how it is modified and ossified over time. In Section~\ref{sec:bar_structure}, we show how this phase skeleton is reflected in the structure of the BP/X bulge in ordinary space. For the first time, we find the position of separatrix orbits in ordinary space and show that the X-structure lies along the separatrix separating near-resonant orbits from those that have undergone transformation and circulate with decreasing resonant angle. In Section~\ref{sec:discussion_conclusions}, we summarize our results.

\section{Numerical model}
\label{sec:model}

We use the numerical model from~\cite{Zozulia_etal2024b,Zozulia_etal2025}, initially consisting of a pure exponential stellar disk and an isotropic spherical dark halo that has an NFW-like density profile~\citep{NFW}. The natural system of units is used with the gravitational constant, $G=1$, the mass of the disk $M_\mathrm{d}=1$, and the initial exponential scale of the disk $R_\mathrm{d}=1$. The initial thickness of the disk is $z_\mathrm{d}=0.05$, where $z_\mathrm{d}$ enters in the ``sech$^2$'' law, the dark halo mass is $M_\mathrm{h}(R<4R_\mathrm{d})\approx1.5$, and the minimum value of the Toomre parameter is $Q(2R_\mathrm{d})=1.2$. 
The components evolve self-consistently until $t=600$ ($\sim8$~Gyr). A fast bar forms by $t\approx50$ and undergoes buckling, reaching maximum vertical asymmetry at $t=180$. From $t=300$, the bar gradually becomes vertically symmetric and grows alongside the BP/X bulge, ending up with the ratio of the bar semi-major axis to the bulge radial extent $a_\mathrm{bar}/r_\mathrm{B{P/X}}\sim3$. The growth is fueled by capturing new disk orbits with higher $H_j$ and lifting them out of the disk mid-plane~\citep{Zozulia_etal2024a,Zozulia_etal2024b,Zozulia_etal2025}.

\par
For the bar and the BP/X bulge, we constructed the evolution of an orbital skeleton consisting of some strictly periodic orbits. To do this, we froze the potential at a given time and calculated it using the AGAMA package \citep{agama}. The disk potential was obtained by expansion in cylindrical harmonics and a dark halo potential by expansion in spherical harmonics. Then, over the entire permissible range of $H_j$ (for bar orbits
with $\Omega-\Omega_\mathrm{p} = \kappa/2$), we found all admissible x1, banana-shaped, and eight-shaped orbits, i.e., x1v1 and x1v2 \citep{Skokos_etal2002a}, as well as orbits with frequency ratios of $\omega_z/\kappa = 5/6$ (brezels, introduced in \citealp{Portail_etal2015b}) and 3/4 (fish-like orbits). The search algorithm for these orbits is described in Appendix~\ref{sec:appendix_b}.

They reflect the general topology of the phase space over the separatrix for x1v1 orbits and divide the space into regions with different frequency ratios, at least for orbits aligned along the major axis of the bar. The fish were chosen because they are strictly periodic orbits, and in our model the distribution of bar orbits by frequency ratio $\omega_z/\kappa$ shows that there are practically no orbits with $\omega_z/\kappa<3/4$ (see figure~7 in \citealp{Parul_etal2020}). For all orbits, we defined the medium-term averaged actions and angles using the \texttt{galport} package \citep{Zozulia_2026}. Further, under actions and angles we always considered medium-term values.

\section{Periodic orbits and phase-space skeleton through the ages}
\label{sec:orbits}

\begin{figure}
\centering
\includegraphics[width=1.0\linewidth]{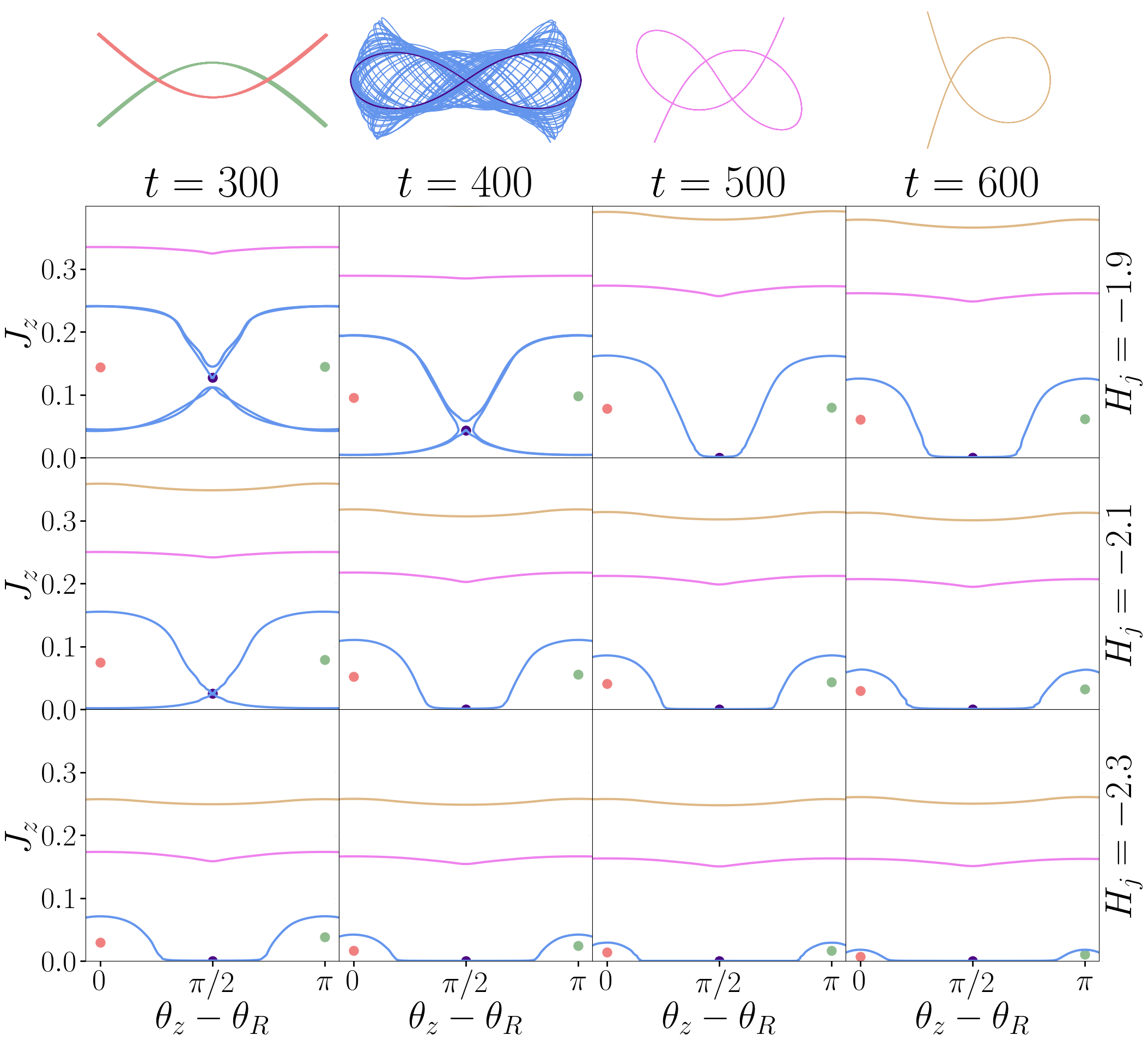}
\caption{Skeleton of the phase space $J_z$--$\theta_\mathrm{res}$. The top row shows examples of orbits corresponding to phase trajectories. \textit{Red} and \textit{green} points represent x1v1 orbits, \textit{blue} points x1v2 orbits, \textit{blue} curves the separatrix, \textit{magenta} brezels, and \textit{yellow} fish. Each subpanel corresponds to a particular Jacobi energy, $H_j$, and time, $t$. 
} 
\label{fig:skeleton}
\end{figure}

In the phase space of actions and angles, x1v1 orbits have a fixed angle, $\theta_\mathrm{res} = \theta_z - \theta_\mathrm{R}=0$\footnote{The angles $\theta_z=0$ and $\theta_R=0$ correspond to the $z$-maxima and apocentre.} (BAN up) or $\pi$ (BAN down), and a fixed value of $J_z$ (at each time moment) for a given $H_j$, i.e., they represent fixed points in the plane $J_z$--$\theta_\mathrm{res}$ (Fig.~\ref{fig:skeleton}, red and green points). Although $H_j$ generally changes as the bar's pattern speed evolves, these changes are small in the central regions. The adiabatic invariant $J_\mathrm{v}=J_\mathrm{R}+L_z/2+J_z$ is conserved~\citep{Zozulia_etal2024a}, giving an approximately one-to-one correspondence between $H_j$ and $J_\mathrm{v}$ for x1 and x1v1 orbits (Appendix~\ref{sec:appendix_a}). Thus, sequences of orbits with the same $H_j$ at different times approximately trace possible orbital transformations.

\par
The orbits x1v2 also have a fixed value of $J_z$ for a given $H_j$ and the angles $\theta_\mathrm{res}=\pi/2$ or $3\pi/2$ (Fig.~\ref{fig:skeleton}, blue points). They are unstable (e.g., \citealp{Skokos_etal2002a}). In the plane $J_z$--$\theta_\mathrm{res}$, they show the points where two branches of the separatrix intersect (Fig.~\ref{fig:skeleton}, blue points), separating near-resonant orbits, librating around fixed points with $\theta_\mathrm{res}=0$ or $\pi$, from two types of circulating orbits: those with small values of $J_z$ and an increasing $\theta_\mathrm{res}$, and those with large values of $J_z$ and decreasing $\theta_\mathrm{res}$ \citep{Zozulia_etal2024b,Zozulia_etal2025}. We describe how the separatrix itself was found in Appendix~\ref{sec:appendix_b}. Remaining periodic orbits (brezels and fish) lie above the separatrix and circulate, decreasing the resonant angle, $\theta_\mathrm{res}=\theta_z - \theta_\mathrm{R}$. We focus only on the separatrix for x1v1 orbits, and do not consider separatrices for other periodic orbits, since the 2:1 resonance is the broadest.
\par
Having found periodic orbits, we constructed a phase-space skeleton from these orbital ``bones''.
Phase portraits of found periodic orbits in the plane $J_z$--$\theta_\mathrm{res}$ were constructed for three values of $H_j=-2.3,\,-2.1,\,-1.9$ at four time moments, $t=300,\,400,\,500,\,600$ (Fig.~\ref{fig:skeleton}). 
It should be noted that $J_z^\mathrm{max}$ for the separatrix at resonant angles 0 and $\pi$ is approximately equal to $2J_z$ for fixed points, as was shown in the model by \citet{Quillen_etal2014}. Fig.~\ref{fig:skeleton} also shows the phase trajectories of the brezels (magenta curves) and fish (yellow curves). 
The shape of the orbits, with phase trajectories presented in Fig.~\ref{fig:skeleton}, is shown in the top row of the figure.

Fig.~\ref{fig:skeleton} shows the evolution of the phase space geometry and topology. For $H_j=-1.9$ at $t=300$, a significant portion of the phase volume (extension along $J_z$) in the region where the fixed points are located is occupied by near-resonant librating orbits located between two branches of the separatrix. By $t=500,\,600$, this region significantly narrows (along $J_z$) and drops downward. The separatrix also shrinks and sinks, which was predicted in a simple Hamiltonian model by \citet{Quillen_etal2014}. As the libration region sinks, it transports particles from high $J_z$ to lower $J_z$, while circulating orbits increase their $J_z$ and move from a region of decreasing resonant angle (below the separatrix) to a region of increasing resonant angle (above the separatrix). This process resembles the levitation mechanism described by \citet{Sridhar_Touma_1996}, but reversed in time. As the libration region shrinks, orbits begin to leave it (see Fig.~\ref{fig:PP_dens_evol} in Appendix~\ref{sec:appendix_c}). 

\par
Transformation occurs more rapidly in central regions, for example, for orbits with $H_j=-2.3$. From $t=300$ to $t=600$, the potential in the central regions of the galaxy hardly changes, so all regular orbits remain in place and the skeleton of the phase space ossifies. In the post-buckling stage (after $t=300$), there are virtually no orbits lying below the lower branch of the separatrix. The region where near-resonant orbits could be located has substantially shrunk. Most of the orbits have already transformed and are located above the separatrix, where brezels and fish circulate. In this case, the fish actually set the maximal value of $J_z$, 
i.e., the ``ceiling'' of the phase space. When $H_j=-2.3$ this boundary remains unchanged starting from $t=300$. For $H_j=-1.9$, from $t=300$ to $t=600$, it continues to slide downward, although the sinking of fish occurs more slowly than the sinking and shrinking of the separatrix. For intermediate values of $H_j$, even at late moments of time, $J_z$ for fixed points with $\theta_\mathrm{res}=0$ and $\pi$ is different from zero, which leaves room in the phase space for near-resonant orbits and their possible subsequent transformation into orbits circulating with a decreasing resonant angle, $\theta_\mathrm{res}$, i.e., lying above the separatrix. 

\par
\begin{figure}
\centering
\includegraphics[width=0.82\linewidth]{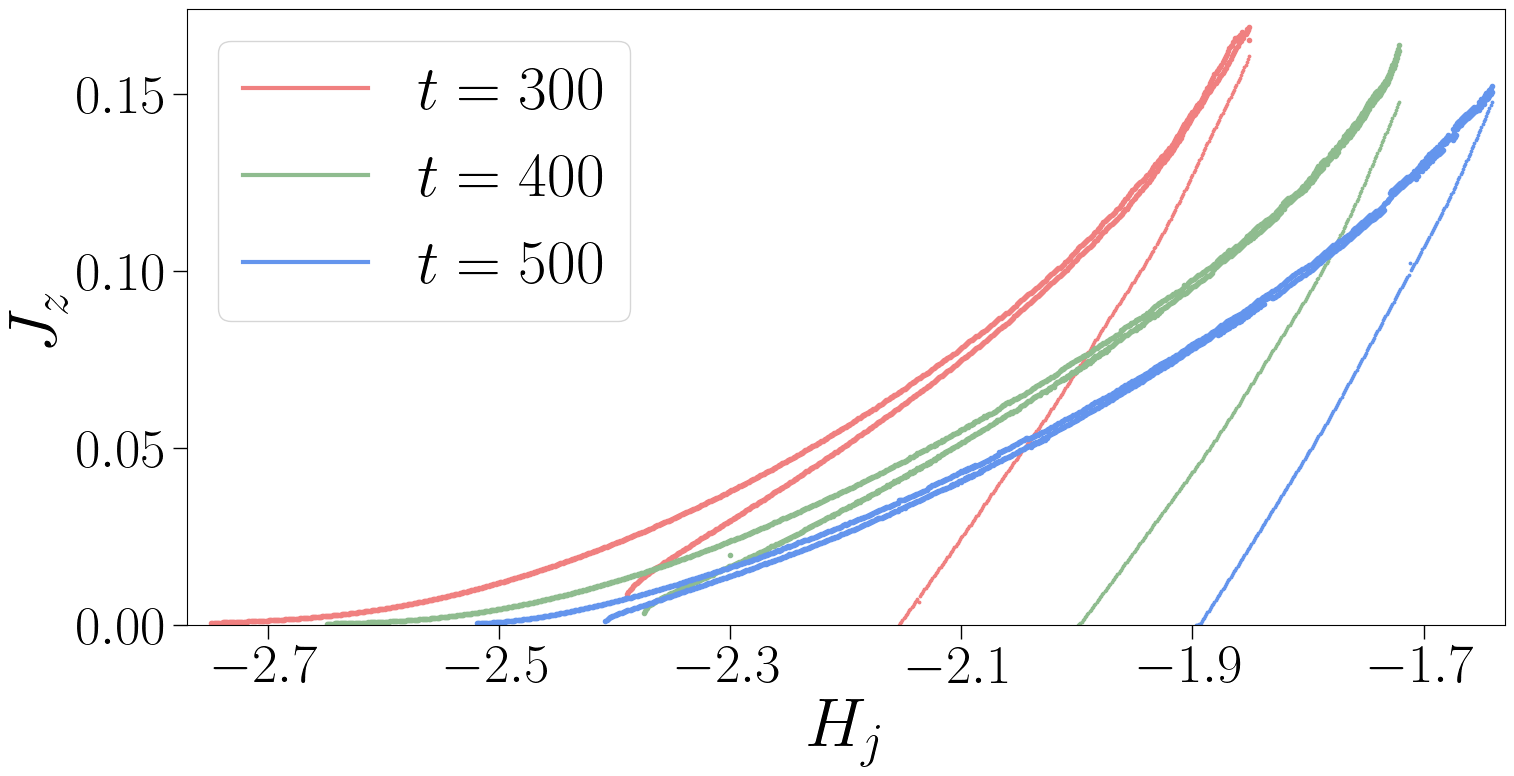}
\caption{Values of $J_z$ versus $H_j$ at different $t$. \textit{Thick} lines: fixed points (x1v1 orbits); $\theta_\mathrm{res}=0,\,\pi$. \textit{Thin} lines: x1v2 orbits; $\theta_\mathrm{res}=\pi/2$.
}
\label{fig:fixed_Hj}
\end{figure}

Fig.~\ref{fig:fixed_Hj} (thick lines) shows the dependence of $J_z$ on $H_j$ for x1v1 orbits at $t=300,\,400,\,500$. 
At $t=300$ the potential still retains asymmetry in the vertical direction, so the sequences of the BAN up and BAN down orbits are different. Such an asymmetry is a legacy of buckling. But this only affects the orbits in the most central regions (small values of $H_j$). Here the asymmetry persists for a long time (see also \citealp{Quillen_etal2014}).  
Orbits x1v1 branch from the x1 family starting from a certain $H_j$, and the later in time that the orbital family is defined, the higher the value of $H_j$ at which the branching occurs. We do not find stable x1v1 orbits over the entire range of $H_j$ values over which the x1 orbits exist. Typically, stable x1v1 orbits terminate earlier than x1 orbits.
\par
Figs.~\ref{fig:skeleton}--\ref{fig:fixed_Hj} show orbits that can exist in a frozen potential. For a fixed time moment, orbits x1v1 have larger values of $J_z$ for larger values of $H_j$. The trend is a decrease in $J_z$ with time as the potential changes. Fixed points seem to sink into the phase space. As a result, banana-shaped orbits become flatter (see Fig.~\ref{fig:x1v1}), facilitating the protrusion of other orbits from the mid-plane.

Fig.~\ref{fig:fixed_Hj} (thin lines) also shows the dependence of $J_z$ on $H_j$ for x1v2 orbits for $t=300,\,400,\,500$.
The same tendency is observed for points corresponding to x1v2 orbits as for fixed points corresponding to x1v1 orbits. They sink into phase space over time. For x1v2 orbits the descent to the bottom of the phase space occurs more rapidly than for x1v1 orbits (see also Fig.~\ref{fig:Hj_t}). It is important to note that the descent of x1v2 orbits to $J_z=0$ removes the region with positive circulation, and the descent of x1v1 orbits towards $J_z=0$ shrinks the libration region and, accordingly, removes purely resonant orbits.
\par
The same patterns are observed for brezel and fish orbits as for the x1v1 and x1v2 orbits. For higher values of $H_j$, $J_z^\mathrm{max}$ is higher and decreases over time, but very slowly (Fig.~\ref{fig:evol_Jzmax_brezel}). The difference in the evolution of $J_z^\mathrm{max}$ is that over time, in the central regions, it stabilizes at a value of $J_z^\mathrm{max}>0$. Fish determine the maximal vertical extension of the BP/X bulge in the central regions, which stops changing over time, and quite early, leaving the BP/X bulge thickness in the central regions unchanged.

\section{General structure of BP/X bulges and its evolution in the ordinary space}
\label{sec:bar_structure}

\begin{figure}
\centering
\includegraphics[width=1\linewidth]{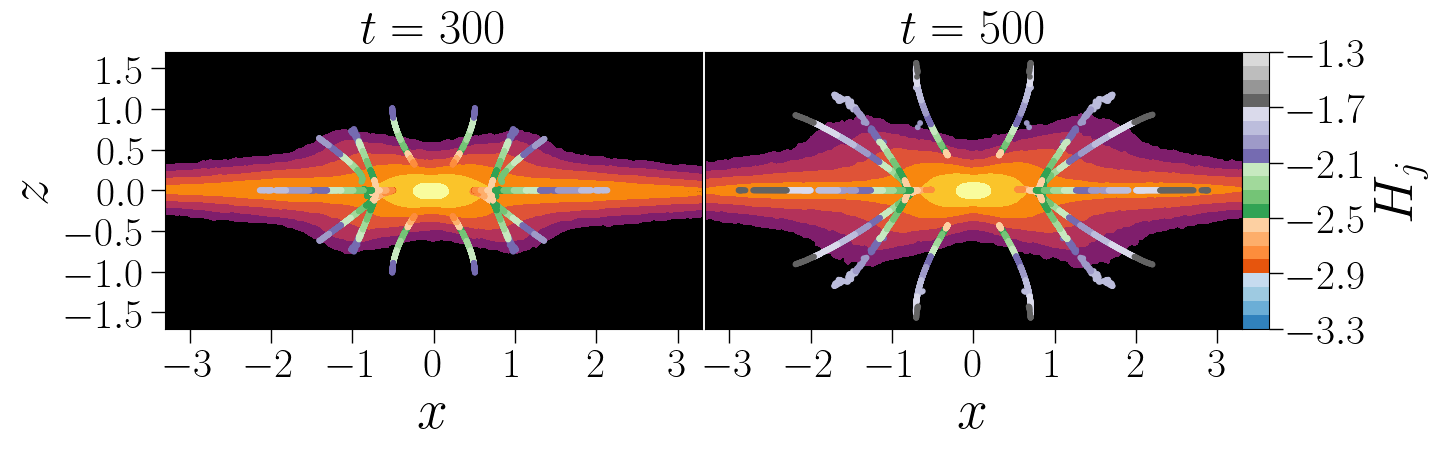}
\caption{$x-z$ snapshot for $t=300$\,(\textit{left}) and $t=500$\,(\textit{right}). 
Individual curves show the locations of the $z$-maxima of the various orbital families and the separatrix of the $x1v1$ family, colored by their $H_j$ values (see the main text).}

\label{fig:xstructure500}
\end{figure}

The phase skeleton in Fig.~\ref{fig:skeleton} can be translated into the orbital structure of the BP/X bulge in ordinary space. For each value of $H_j$, using the ~\texttt{AGAMA}~\citep{agama} package, for all types of periodic orbits studied, we associate $J_z^\mathrm{max}$ with $(x,z_\mathrm{max})$. The points $(x,z_\mathrm{max})$ for the main types of periodic orbits at $t=300$ and $t=500$ are shown in Fig.~\ref{fig:xstructure500} (left and right panels, respectively), overlaid on model snapshots in the $(x,z)$ plane, with colors indicating surface density. 
It can be seen that the curves for the x1v1 orbits (fixed points in phase space) with the largest values of $H_j$ lie on the very periphery of the BP/X bulge, not in its thickest part.
The fish fall on its outer border (funnel) in the central part. The curve for separatrix orbits originates from the highest point of the X-structure of the BP/X bulge.
And this is true for any moment in time. In fact, the outermost parts of the X-structure rays are outlined by orbits lying under the separatrix. When the separatrix descends and reaches a region with a large number of particles, orbits begin to be captured into resonance near it and then pass through it. Therefore, particles with the largest $H_j$, recently captured by the bar, rise and outline the rays of the X-structure at its very end. There is a crowding of orbits here. We illustrate the phenomenon of orbit accumulation near the separatrix in Appendix~\ref{sec:appendix_c}. The separatrix with the largest value of $H_j$ and the orbits adjacent to it determine not only the beginning of the X-structure ray, but also the ray itself (Fig.~\ref{fig:sep_orbit}). Thus, the separatrix acts as the armature of the BP/X bulge. A connection can be traced between the orbits adjacent to the separatrix (Fig.~\ref{fig:PP_dens_evol}) and the orbits associated with unstable x1v2 family that remain ``sticky'' to the tori surrounding stable x1v1 orbits at a given $H_j$ (\citealp{Patsis_Harsoula2018}, figure~8).
The central part of the BP/X bulge is assembled from orbits above the separatrix, spanning a wide range of heights at a given $H_j$.

\par
All curves in Fig.~\ref{fig:xstructure500} are colored according to the values of $H_j$, and the transition along the same value of $H_j$ between different types of orbits shows the direction of their transformation --- from x1v1 (or near-resonant orbits) through the separatrix to brezels (and orbits around them) and then to fish. An important point should be made. The sequence of admissible fish orbits at $t=500$ extends to the same $H_j$ values as the sequence of x1v1 orbits. In Fig.~\ref{fig:xstructure500}, the outermost drawn isophote corresponds to fish orbits with much lower $J_\mathrm{v}$ values. Beyond this isophote, there are no fish orbits, i.e., the corresponding bones have not had time to develop ``muscle'' growth. The orbits may appear much later, when sufficient time has passed for the x1v1 orbits to transform and the separatrix at large values of $H_j$ drops significantly to the bottom of the phase space. At $t=500$, the last fish orbits actually found in the $N$-body model correspond to x1v1 orbits captured by the bar much earlier.
\par
Unstable x1v2 orbits define the separatrix, which forms the ``backbone'' of the entire orbital structure. Other periodic orbits are attached as bones to it. Together, they form the orbital ``skeleton'' of the  BP/X bulge.
The skeleton itself transforms over time. All parts of the skeleton shrink with time along $J_z$ (Fig.~\ref{fig:skeleton}). The same shrinkage is visible in snapshots (Fig.~\ref{fig:xstructure500}). Shrinkage occurs more rapidly in the central regions. There, the skeleton ossifies by $t=300$, while at the periphery, it is mobile even at $t=600$.

\section{Summary}
\label{sec:discussion_conclusions}

We investigate the phase-space evolution of periodic-orbit families in the BP/X bulge, using Cartesian and action-angle spaces. Bar-aligned 3D orbits with low frequency ratios lie behind the separatrix. Below its upper branch, we distinguish near-resonant orbits and orbits that have not yet undergone transformation and circulate with an increasing resonant angle.
\par
Initially, when the separatrix surrounds fixed points at $\theta_\mathrm{res}=0$ and $\pi$ on both vertical action sides, many low-$J_z$ circulating orbits can be resonantly captured or passed. Later, as the separatrix region shrinks and sinks to the bottom of the phase space, only resonant heating occurs. This process varies across the bar. Resonant capture uplifts orbits fastest in the center, where x1v1 orbits are flat, and almost all orbits lie above the separatrix. They form the central BP/X bulge. Meanwhile, in intermediate and peripheral regions, all orbit types briefly coexist.
\par
The phase space evolution patterns we discovered are generally similar to those reported in \citet{Quillen_etal2014}, who studied them using phenomenological considerations and a simple Hamiltonian model. Here we have traced the evolution of 
an $N$-body model in phase space over a long timescale. 

We have taken the next important step in understanding the orbital structure of BP/X bulges and their long-term evolution. It turned out that not only does the separatrix shrink and sink, but also all the bones of the phase skeleton shrink until the skeleton ossifies. At this point, the separatrix and the fixed points sink to the bottom of the phase space. In different areas of the BP/X bulge, this happens at different rates.
\par
The phase volumes occupied by different types of orbit gradually change, but they are separated by one feature -- a separatrix. This feature also manifests itself in the ordinary $xz$ space and can be revealed in snapshots as the uppermost end of the X-structure. 
We found that it is precisely the separatrix orbits associated with unstable figure-of-eight orbits that give the 3D bar its characteristic peanut/X-shaped appearance. These orbits define the skeleton of the entire 3D structure and the outer contour of the 3D bar follows the contour of the separatrix orbit (Fig.~\ref{fig:skeleton}) with the largest value of the Jacobi integral for orbits that began to rise out of the plane. Here, the transformation of orbits is currently ongoing. Other orbits previously proposed in the literature as building blocks of the BP/X bulge are the filling of the region within this contour. They lie beyond the separatrix and, although exhibiting a peanut/X-shaped morphology (but only in unsharp-masked images), are located in the very central regions. They rose a long time ago and not very high, and their transformation is almost complete. As for the orbits trapped in the resonance associated with the x1v1 orbits, they build up the BP/X bulge in its peripheral and still-thin region.
\par
As the separatrix descends in phase space, $z_\mathrm{max}$ corresponding to its $J_z^\mathrm{max}$ decreases. Consequently, the opening angle of the X-structure decreases too, while the rays become longer. This occurs due to the addition of new orbits with larger $H_j$ to the bar and the BP/X bulge. This evolution of the opening angle is precisely observed in $N$-body models \citep{Smirnov_Sotnikova2018}.

\begin{acknowledgements}

We thank the anonymous reviewer for the valuable suggestions.
We acknowledge financial support from the Russian Science Foundation, grant no. 25-22-00738.
\par
We acknowledge the use of the~\texttt{AGAMA}~\citep{agama}  \texttt{Python} package, without which the present work would not be possible.

\end{acknowledgements}

\bibliographystyle{mnras}
\bibliography{aa62841-26}

\begin{appendix}
\section{Jacobi integral and adiabatic invariant}
\label{sec:appendix_a}

\begin{figure}[H]
\centering
\includegraphics[width=1.0\linewidth]{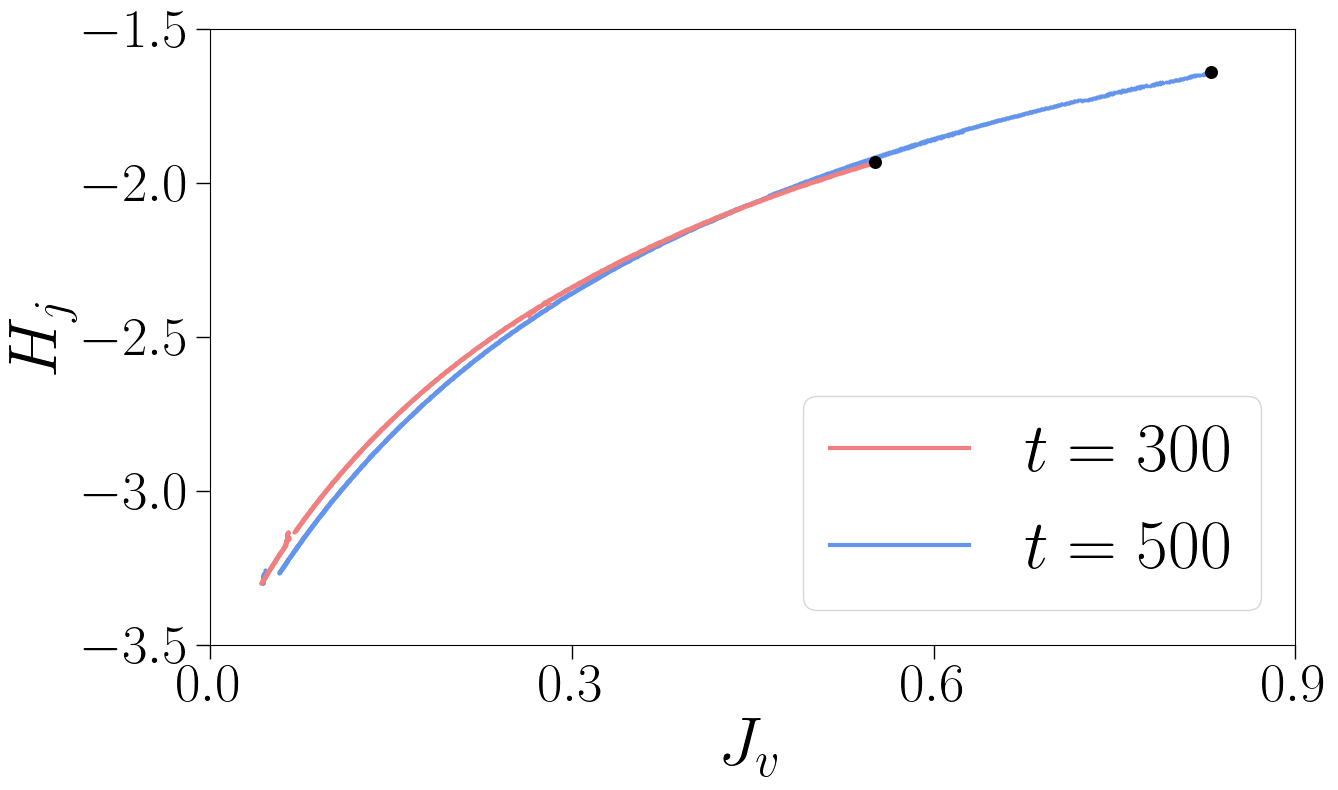}
\caption{$H_j$ versus $J_\mathrm{v}$ for x1v1 orbits. \textit{Red} line --- $t=300$, \textit{blue} line --- $t=500$. \textit{Black} dots are the boundaries for x1 orbits.}
\label{fig:Hj_J_v}
\end{figure}

According to \citet{Zozulia_etal2024a}, for orbits satisfying the relation $2(\Omega - \Omega_\mathrm{p}) \approx \kappa$ and $\omega_x \approx \omega_z$, the adiabatic invariant $J_\mathrm{v}=J_\mathrm{R}+J_z+L_\mathrm{z}/2$ is conserved.
Figure~\ref{fig:Hj_J_v} shows that the Jacobi integral $H_j$ is practically preserved for x1v1 orbits, as is the value of the adiabatic invariant $J_\mathrm{v}$, indicating a one-to-one relationship between them. Boundaries of $H_j$ and $J_\mathrm{v}$ for x1 orbits, from which x1v1 orbits branch, at different times are marked by black dots.

\section{Algorithm for searching periodic orbits in an $N$-body potential}
\label{sec:appendix_b}

\textit{Orbits x1 and x1v1.} We determine the pattern speed $\Omega_\mathrm{p}(t)$ of a bar, then we freeze a potential $\Phi$ of our $N$-body model at a given time moment $t$.  We search periodic orbits in 
such a frozen potential for a fixed value of $H_j = (v_x^2 + v_y^2 + v_z^2)/2 + \Phi(x,y,z) - \Omega_\mathrm{p}\cdot(x\cdot v_y - y\cdot v_x)$. All other periodic orbits, except for the x1v1 orbits, are searched for in a symmetrized potential. This is calculated automatically using the AGAMA package. Triaxial symmetry is used. The search for x1 orbits was carried out at $y=z=v_x=v_z=0$, $x$ was varied, $v_y$ was found from $H_j$ and other phase coordinates. Then we minimize the quantity $\sum |\mathbf{r}_\mathrm{apo,i} - \mathbf{r}_\mathrm{apo,0}|^2$
in the apocenter at $y=0$, $x>0$ every fourth moment when 
$v_\mathrm{R}=x\cdot v_x+y\cdot v_y=0$.
\par
As stated above, the orbits of x1v1 are sought in an asymmetric potential. To find the first orbit of the x1v1 family, we choose and fix $H_j=-2.3$. For such a value of $H_j$, an orbit definitely exists. A fixed point is sought in the phase space on the set $y=0,\,v_x=0,\,v_z=0$, which lies on the phase trajectory corresponding to the orbit x1v1. The coordinates $x$ and $z$ are varied, and $v_y$ is calculated from $H_j$, $x$, and $z$. We look for $x$ and $z$ that correspond to the minimum of the sum of the squares of the residuals at the apocenter $y=0$, $x>0$ every fourth moment when $v_\mathrm{R}=0$. After finding the ``base'' closed x1v1 orbit, we begin to vary $H_j$. The coordinates $x$ and $z$ obtained in the previous step (for the previous $H_j$) are used as the initial conditions for $x$ and $z$ for the next orbit. The search proceeds in both directions: toward decreasing $H_j$ and toward increasing $H_j$.
\par
\begin{figure}
\centering
\includegraphics[width=1\linewidth]{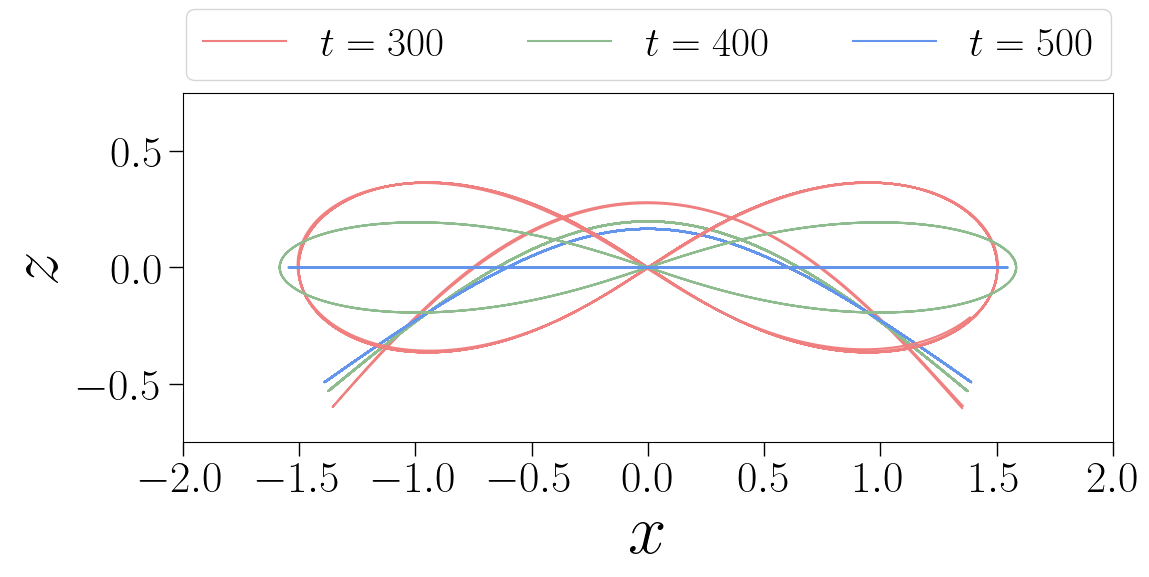}
\caption{Orbits x1v1 and x1v2 with the same $H_j$ at three time moments in a frozen potential, $xz$ projections. Orbits x1v1 were determined in an asymmetric potential. \textit{Red} line $t=300$,  \textit{green} line $t=400$, \textit{blue} line $t=500$; $H_j=-1.95$.}
\label{fig:x1v1} 
\end{figure}

An example of a pure banana-shaped orbit for $H_j=-1.95$ in the frozen potential is shown in Fig.~\ref{fig:x1v1}. One can see how the orbit flattens over time, from $t=300$ to $t=500$.

\begin{figure}
\centering
\includegraphics[width=1.0\linewidth]{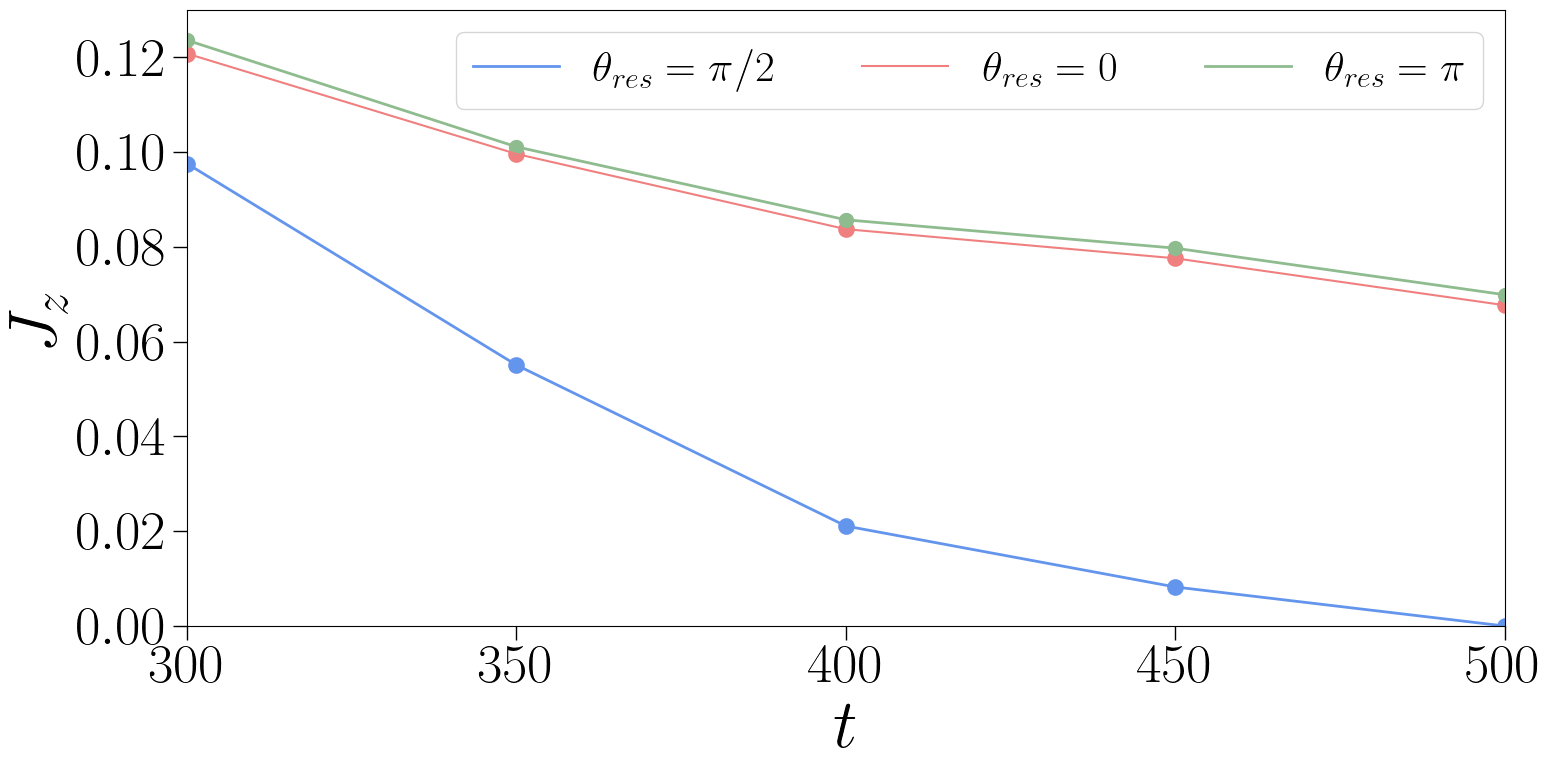}
\caption{Evolution of the vertical action $J_z$ for fixed points (x1v1 and x1v2 orbits). $H_j=-1.95$.
}
\label{fig:Hj_t}
\end{figure}

\textit{Separatrix and x1v2 orbits.} Orbits x1v2 (eight-shaped) branch from orbits x1 at large values of $H_j$, therefore the first orbit of this family is sought in the region of large $H_j$. The starting orbit is the orbit x1. We fix the values of $H_j$, $y=0$, and $z=0$, $v_x=0$, $x$ is varied. The start is initiated with $v_{z,\,\mathrm{ini}}=0$. Then $v_z$ increases with a step of 0.001, and $v_y$ is calculated from $H_j$, $x$ and $v_z$. For each value of $v_z$, we seek $x$ that minimizes the sum of squared residuals at the apocenter ($y=0$, $z>0$) at every fourth instant, when $v_\mathrm{R}=0$. Thus, a grid-specified dependence of the sum of squared residuals on $v_z$ is obtained. The value of $v_z$ is taken at which the sum of squared residuals has a minimum (all cases of $v_z=0$, or close to 0 are rejected). Then, the $v_z$ variation is turned on and a more precise eight-shaped orbit is found. For the next value of $H_j$, $x$, and $v_z$ from the previous step are used as initial conditions. The search proceeds in both directions: toward decreasing $H_j$ and toward increasing $H_j$. The examples of x1v2 orbits with $H_j=-1.95$ in a frozen symmetrized potential at $t=300$, $t=400$ and $t=500$ are shown in Fig.~\ref{fig:x1v1}. The change in $J_z$ with time for a given $H_j$ for orbits x1v1 and x1v2 is shown in Fig.~\ref{fig:Hj_t}.
\par
The separatrix is obtained from the orbit x1v2 with a given value of $H_j$. The orbit x1v2 is unstable and, if integrated over a long time interval ($\Delta t \sim 300$), emerges from the point $(J_z,\theta_\mathrm{res}=\pi/2)$ and traces a separatrix in the plane $J_z$--$\theta_\mathrm{res}$. For small values of $H_j$ (at which x1v1 are not yet degenerate, but x1v2 are already degenerate), this procedure is not suitable. A manual approach is used here. We start from the banana-shaped orbit at the corresponding $H_j$. $z$ is fixed at the apocenter, $x$ is varied, and $v_y$ is found from $H_j$ and the remaining phase coordinates. We gradually increase $z$ by 0.001. The sum of the squared residuals at every 4th moment when $v_\mathrm{R} = 0$ is minimized. The procedure continues until the banana orbit 
unwinds so much that the libration amplitude along $J_z$ reaches its maximum. Then the position of the separatrix in the plane $J_z$--$\theta_\mathrm{res}$ is found by smoothing the phase trajectory of the perturbed x1v2 orbit over medium-term timescales \citep{Zozulia_2026}.

\textit{Brezels and fish.} To find brezel and fish orbits we symmetrized the potential and did not distinguish between orbits with $J_z^\mathrm{max}$ near $\theta_\mathrm{res}=0$ and $\pi$. To find the base brezel orbit, we start from the orbit x1 with a small value of $H_j=-2.5$. Minimization of the residuals in spatial coordinates is performed at every 12th instant when $x \cdot v_x + z \cdot v_z = 0$, i.e., at the highest point of the orbit with $x>0$. After finding the preliminary brezel, we switch on a variation in $z$ to obtain a more accurate orbit. Then, $H_j$ is varied, both increasing and decreasing $H_j$.
\par
\begin{figure}
\centering
\includegraphics[width=0.49\linewidth]{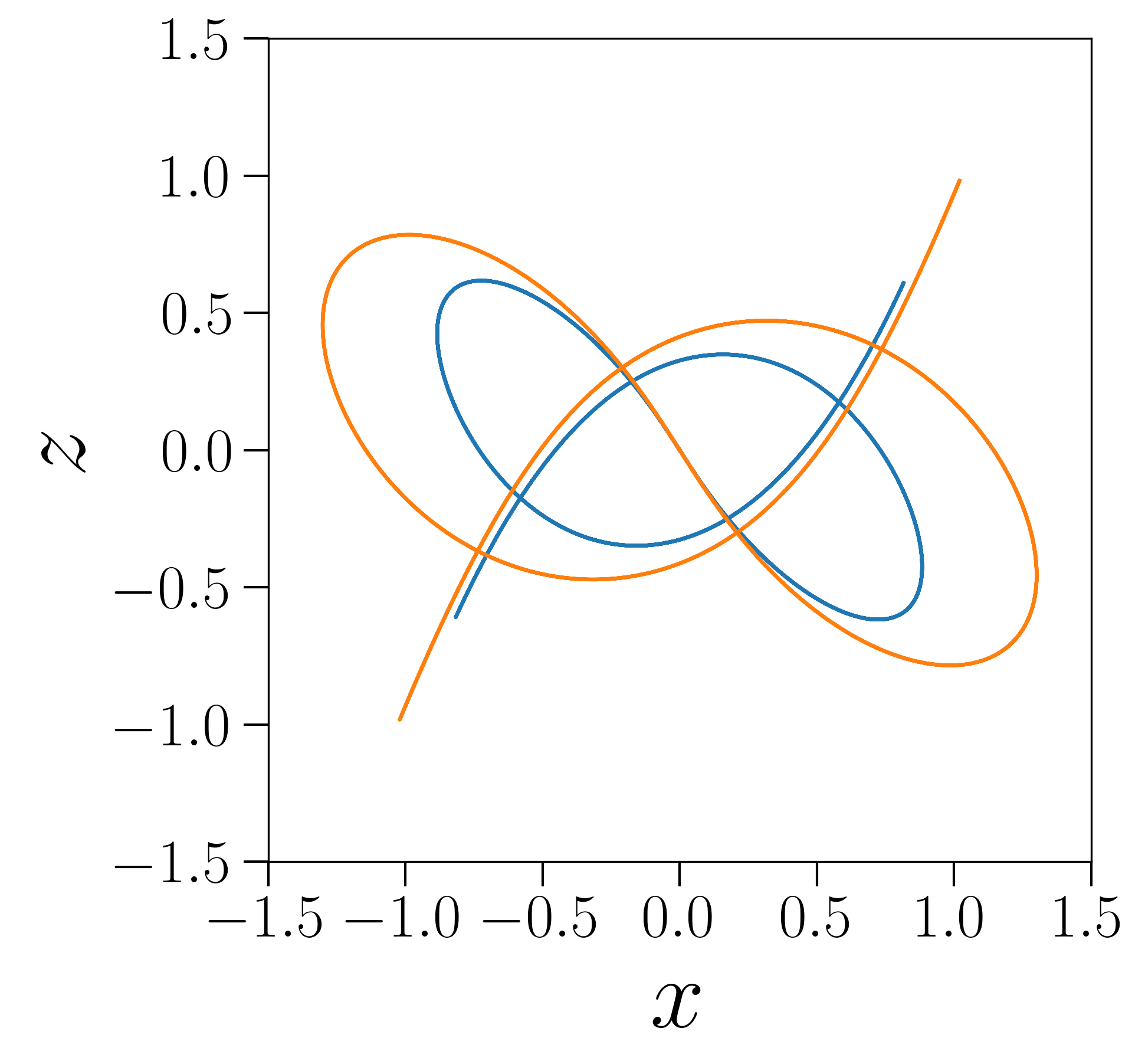}
\includegraphics[width=0.49\linewidth]{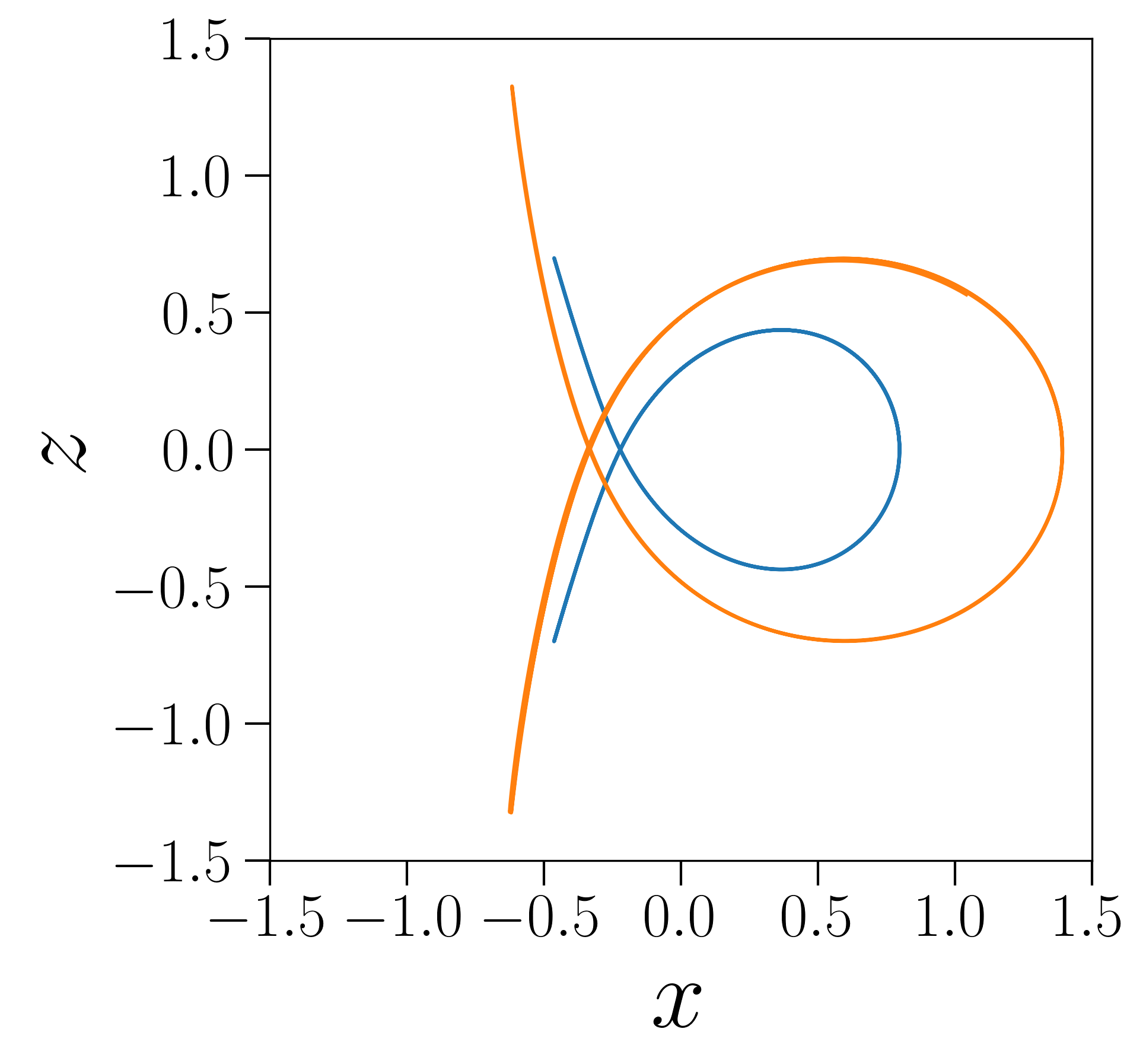}
\caption{Brezel (\textit{left}) and fish (\textit{right}) orbits in a frozen potential at $t=400$. \textit{Blue} line $H_j=-2.2$, \textit{yellow} line $H_j=-1.9$.
}
\label{fig:brezel}
\end{figure}
\begin{figure}
\centering
\includegraphics[width=0.49\linewidth]{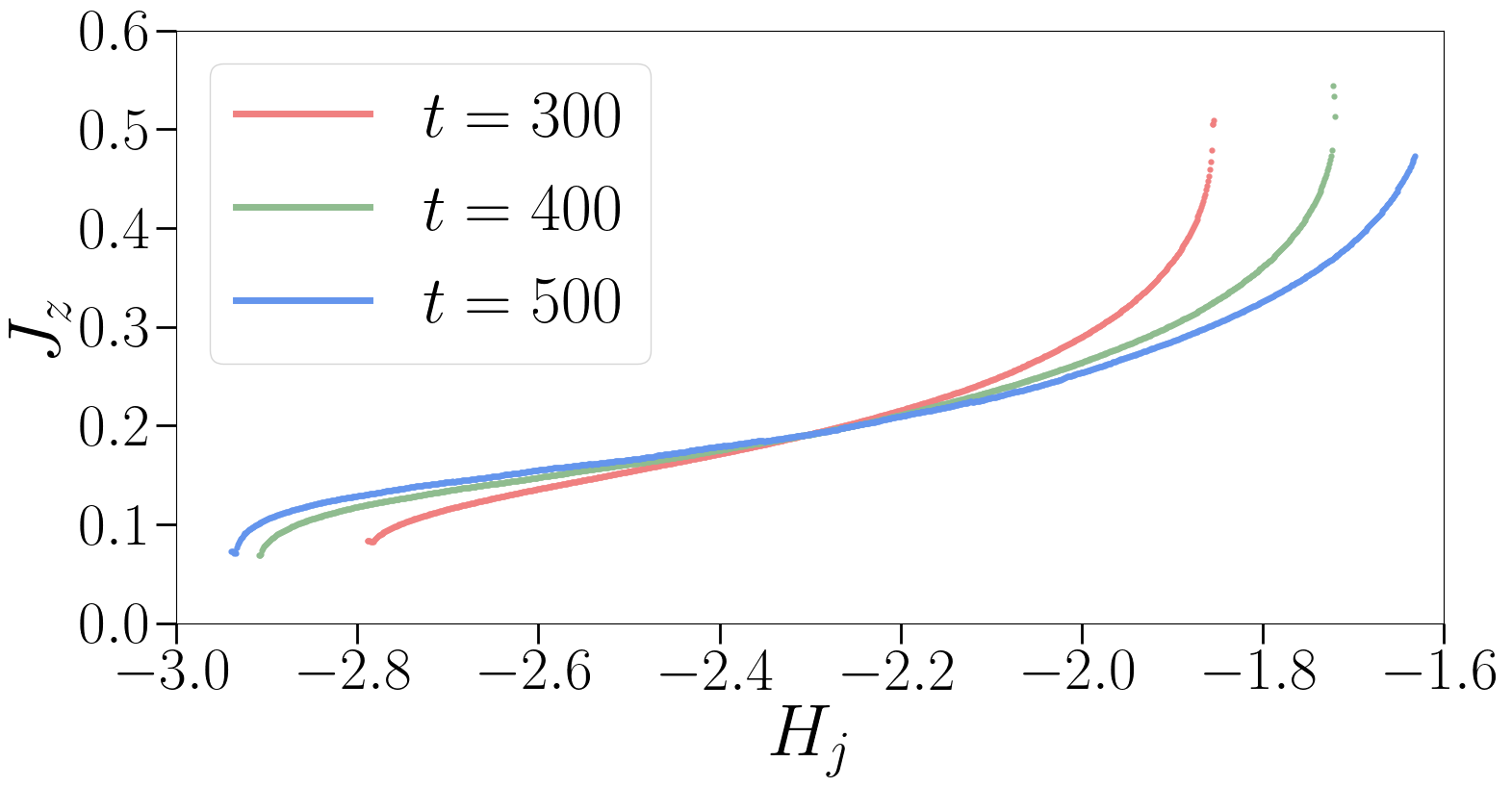}
\includegraphics[width=0.49\linewidth]{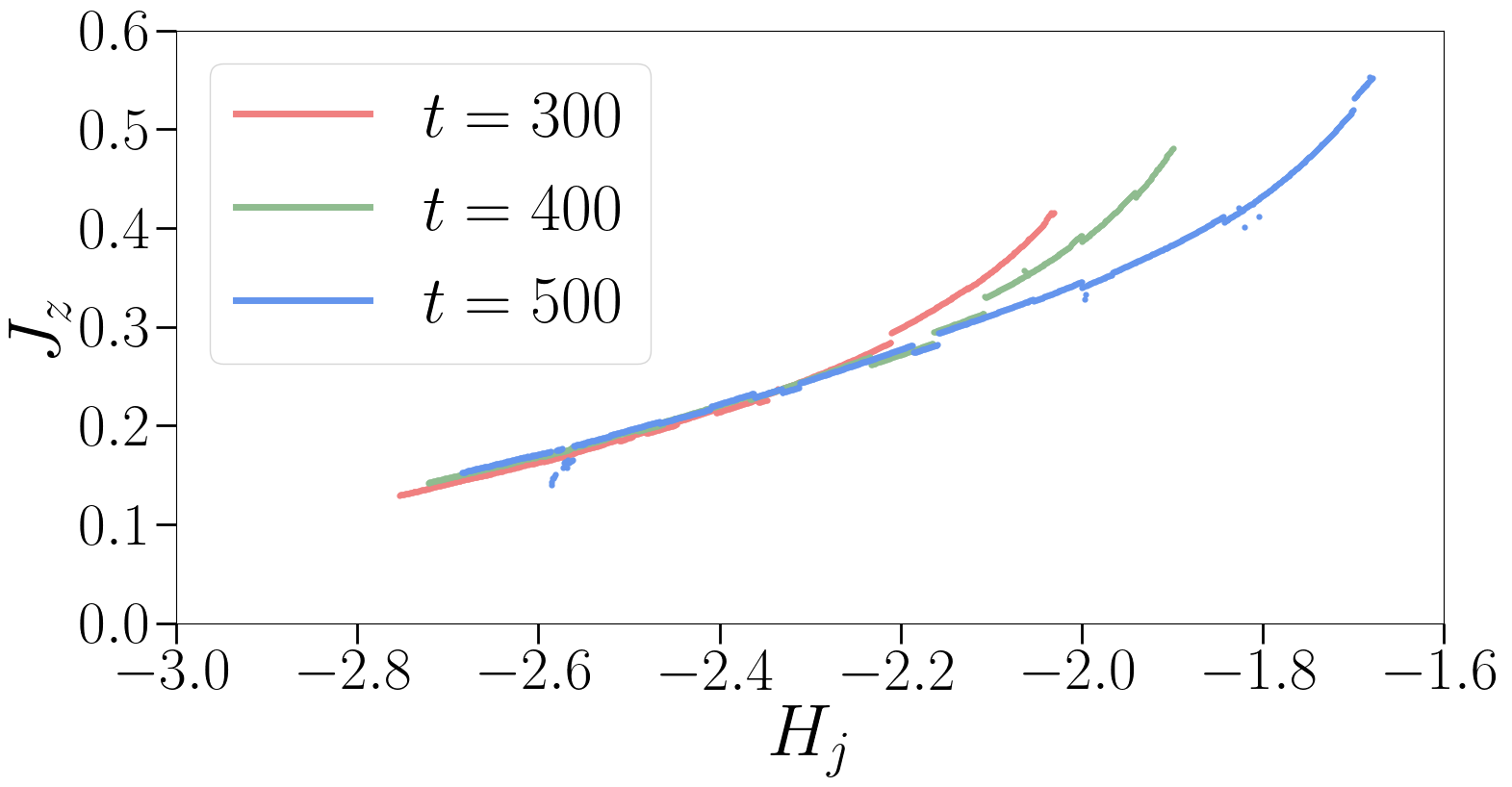}
\caption{The values of $J_z$ versus $H_j$ at different $t$. \textit{Left} plot: brezels; \textit{right} plot: fish; $\theta_\mathrm{res}=0,\,\pi$.
}
\label{fig:evol_Jzmax_brezel}
\end{figure}

The search for fish-shaped orbits is carried out in the same way as the search for orbits x1v2, with the only difference that at the very beginning an orbit x1 with a small value of $H_j=-2.5$ is taken, on which the orbit x1v2 degenerates into a flat orbit x1. Examples of brezel and fish orbits in a frozen symmetrized potential at $t=400$ are shown in Fig.~\ref{fig:brezel} (left and right plots, respectively). The position of brezel and fish orbits in the plane $J_z$--$\theta_\mathrm{res}$ was found by smoothing the phase trajectory over medium-term timescales. Fig.~\ref{fig:evol_Jzmax_brezel} shows the evolution of $J_z^\mathrm{max}$ for brezel (left plot) and fish (right plot) orbits.

\section{Evolution of phase density in action-angle space}
\label{sec:appendix_c}

\begin{figure}
\centering
\includegraphics[width=1\linewidth]{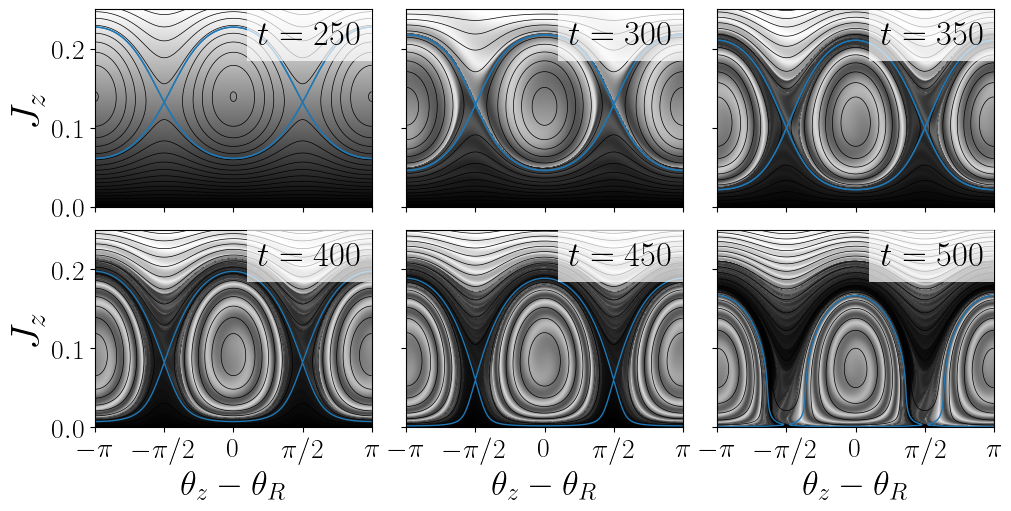}
\caption{Qualitative picture of the phase-space density evolution for bar-aligned 3D orbits with $H_j = -1.9$. Black lines correspond to isolines of the Hamiltonian~\eqref{eq:appendix_ham}, while the blue line denotes the separatrix (derived from the analytical Hamiltonian, not to be confused with the separatrix orbit from Appendix~\ref{sec:appendix_b}).}
\label{fig:PP_dens_evol}
\end{figure}

The qualitative picture of the phase-space density evolution is shown in Fig.~\ref{fig:PP_dens_evol} for bar-aligned 3D orbits with $H_j = -1.9$. This example demonstrates the lowering of the resonance region, as well as the capture and passage of particles through it. Most importantly, it illustrates how particles accumulate near the separatrix.
\par
To trace the density evolution in action-angle space ($\theta_\mathrm{res} = \theta_z - \theta_R, J_z$), we construct an analytical model of the Hamiltonian using the \texttt{galport} package \citep{Zozulia_2026}:
\begin{equation}
\label{eq:appendix_ham}
H(J_z, \theta_\mathrm{res}, t) = \sum_{j=1}^4 \left[ a_j(t) + b_j(t) \cos 2\theta_\mathrm{res} \right] J_z^{0.5j}.
\end{equation}
The coefficients $a_j(t)$ and $b_j(t)$ are computed at time moments from $t = 250$ to $500$ with a step of $\Delta t = 50$ using an optimization algorithm, and are linearly interpolated for intermediate times. Using an 8th-order Runge-Kutta integrator, we integrate a grid of orbits backward in time within this evolving Hamiltonian. By Liouville's theorem, the phase-space density at time $t$ is mapped from the state at $t = 250$, i.e., $f(J_z(t), \theta_\mathrm{res}(t)) = f(J_z(250), \theta_\mathrm{res}(250))$. The initial distribution at $t = 250$ 
is chosen to be linear in $J_z$ and independent of $\theta_\mathrm{res}$ for clarity. The moment $t = 250$ was selected as the starting point because the buckling phase has already ended, while the libration region remains at a sufficiently high $J_z$ for the given $H_j$.

\section{Separatrix orbit in ordinary space}
\label{sec:appendix_d}

\begin{figure}[H]
\centering
\includegraphics[width=1\linewidth]{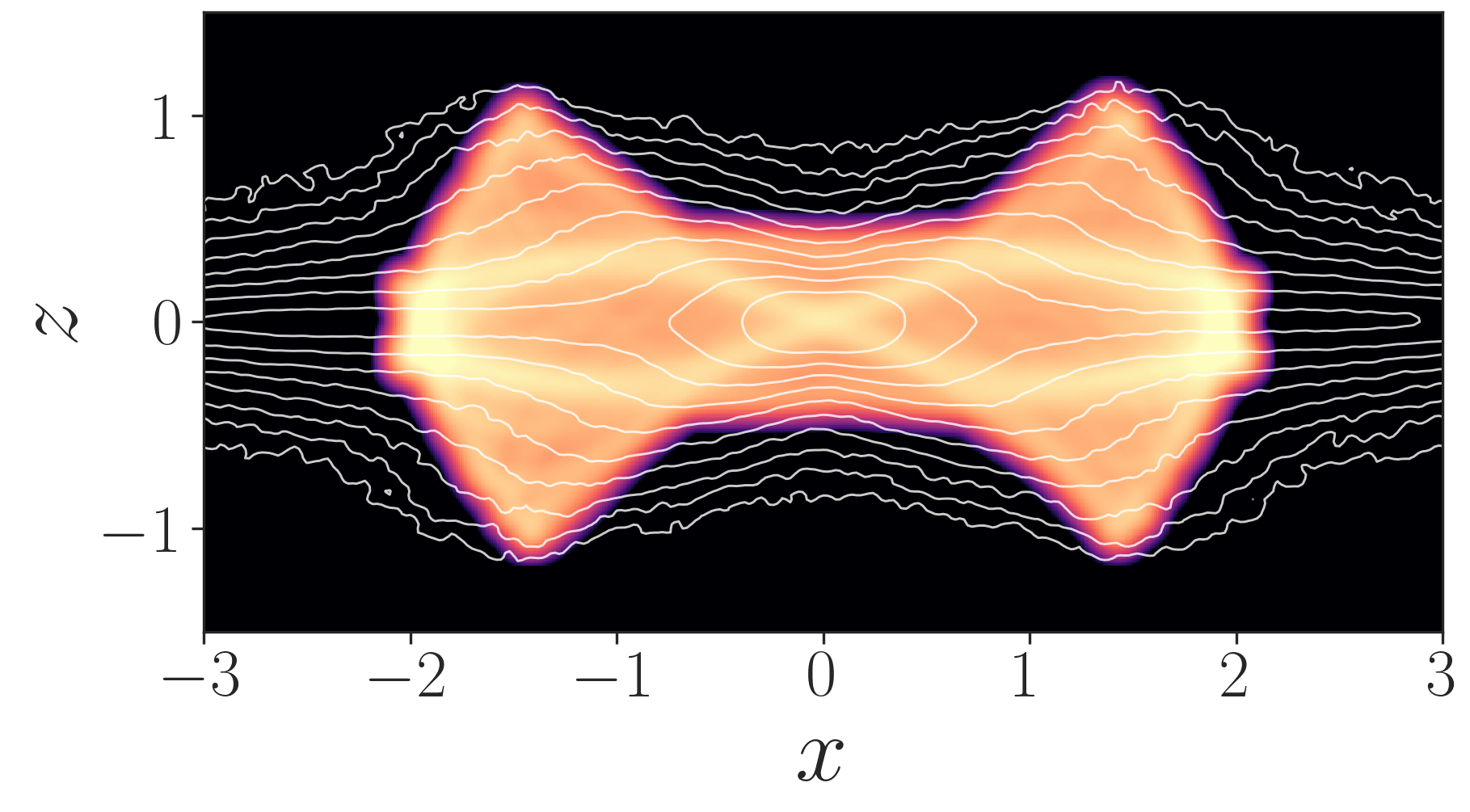}
\caption{Density distribution in the $xz$ plane created by particles on a separatrix orbit with $H_j=-1.8$ at $t=500$. White lines show isophotes of the model at the corresponding time.}
\label{fig:sep_orbit}
\end{figure}

To demonstrate how the separatrix orbit structures the thick part of the BP/X bulge and the ray of the X-structure, we smoothed such an orbit with high $H_j$ by averaging the particle's time in orbit. Fig.~\ref{fig:sep_orbit} demonstrates that the X-structure ray passes through the region containing the caustic, i.e., the area where the orbital loops are concentrated.

\end{appendix}

\label{lastpage}
\end{document}